\documentclass[preprint,12pt,3p]{elsarticle}

\usepackage{graphics}
\usepackage{graphicx}
\usepackage{epsfig}

\usepackage{xcolor}
\usepackage{times}
\usepackage{epsfig}
\usepackage{graphicx}
\usepackage{subfig}
\usepackage{float}
\usepackage{multirow}
\usepackage{setspace}

\usepackage{color}
\usepackage{amsmath}
\usepackage{algorithm}
\usepackage{algorithmic}
\usepackage{bm}
\usepackage{latexsym}
\usepackage{amsthm}
\usepackage{url}
\usepackage{amsfonts}
\usepackage{amssymb}
\usepackage{indentfirst}
\usepackage{float}
\usepackage{tabularray}
\usepackage{booktabs}
\makeatletter \@fpsep = 2pt \makeatother
\begin{document}

\begin{spacing}{1.0}

\begin{frontmatter}

\title{A cross Transformer for image denoising}

\author[label1,label2]{Chunwei Tian\corref{cor1}}
\ead{chunweitian@nwpu.edu.cn}
\author[label1]{Menghua Zheng}
\author[label3,label4]{Wangmeng Zuo}
\author[label5]{Shichao Zhang}
\cortext[cor1]{Corresponding author}
\author[label6,label2]{Yanning Zhang\corref{cor1}}
\ead{ynzhang@nwpu.edu.cn}
\author[label7]{Chia-Wen Lin}

\address[label1]{School of Software, Northwestern Polytechnical University, Xi’an, Shaanxi, 710129, China}
\address[label2]{National Engineering Laboratory for Integrated Aero-Space-Ground-Ocean Big Data Application Technology, Xi’an, Shaanxi, 710129, China}
\address[label3]{School of Computer Science and Technology, Harbin Institute of Technology, Harbin, Heilongjiang, 150001, China}
\address[label4]{Peng Cheng Laboratory, Shenzhen, 518055, China}
\address[label5]{School of Computer Science and Engineering, Central South University, Changsha, Hunan, 410083, China}
\address[label6]{School of Computer Science, Northwestern Polytechnical University, Xi’an, Shaanxi, 710129, China}
\address[label7]{Department of Electrical Engineering and the Institute of Communications Engineering, National Tsing Hua University, Hsinchu, 30013, Taiwan.}

\begin{abstract}
Deep convolutional neural networks (CNNs) depend on feedforward and feedback ways to obtain good performance in image denoising. However, how to obtain effective structural information via CNNs to efficiently represent given noisy images is key for complex scenes. In this paper, we propose a cross Transformer denoising CNN (CTNet) with a serial block (SB), a parallel block (PB), and a residual block (RB) to obtain clean images for complex scenes. A SB uses an enhanced residual architecture to deeply search structural information for image denoising. To avoid loss of key information, PB uses three heterogeneous networks to implement multiple interactions of multi-level features to broadly search for extra information for improving the adaptability of an obtained denoiser for complex scenes. Also, to improve denoising performance, Transformer mechanisms are embedded into the SB and PB to extract complementary salient features for effectively removing noise in terms of pixel relations. Finally, a RB is applied to acquire clean images. Experiments illustrate that our CTNet is superior to some popular denoising methods in terms of real and synthetic image denoising. It is suitable to mobile digital devices, i.e., phones. Codes can be obtained at https://github.com/hellloxiaotian/CTNet. 
\end{abstract}


\end{frontmatter}

% \linenumbers

\section{Introduction}
\label{sec-1}

Image denoising can restore clean images from given corrupted images collected by digital cameras \cite{tian2020deep}, which can provide an important guarantee for high-level computer vision tasks\cite{imani2020overview}, i.e., autonomous driving \cite{levinson2011towards} and identification \cite{bharadwaj2011quality}. Most denoising methods rely on a degradation method \cite{levin2011natural} of ${I_n} = {I_c} + n$ to achieve a denoiser, where $I_n$ is symbolled as a given noisy image, $n$ is defined as noise (i.e., additive white Gaussian noise (AWGN) with variance  $\sigma$) and $I_c$ is used to express a clean image. According to the mentioned illustrations, we can see that the image denoising task is an ill-posed problem. To resolve it, Bayesian theory is referred to remove the noise \cite{levin2011natural}. In terms of mathematics, Markov Random Field (MRF) used geometric prior to smooth captured noisy images for enhancing denoising effect \cite{malfait1997wavelet}. To improve denoising performance, Block-matching and 3D filtering (BM3D) constructed a space mapping from non-local 2D space into a 3D array space to implement a filter for better suppressing noise \cite{dabov2007image}. Dong et al. embedded a sparse representation into a clustering method to obtain different information containing local and non-local information for obtaining high-quality images \cite{dong2011sparsity}. To improve the capability and flexibility of a denoising model, a weighted nuclear norm minimization (WNNM) assigned different weights to obtain singular values, which collaborated with image nonlocal self-similarity to restore high-definition pixel points in image denoising \cite{gu2014weighted}.  

Although these methods have obtained better denoising results in some scenes, they suffer from the following drawbacks \cite{zhang2017beyond}. (1) They pursue optimal parameters through manual tuning parameters. (2) They usually sacrifice efficiency by using complex optimization algorithms to improve quality of repaired images. To address the problems above, discriminative learning methods, i.e., convolutional neural networks (CNNs) with end-to-end architectures act graphics processing unit of powerful computational capabilities are developed in image denoising \cite{zhang2017beyond}.  Zhang et al. utilized common components, i.e., convolutional layer, batch normalization (BN) and Rectified Linear Units (ReLU) to achieve an efficient denoising network in terms of denoising performance and complexity \cite{zhang2017beyond}. To exploit the prior from similar images for complement, Xu et al. \cite{xu2022model} applied the idea of computational graph, deep implicit distribution and deep prior to optimize the image denoiser. To eliminate effect of complex background, Tian et al. exploited an attention mechanism, sparse mechanism and signal processing operations to design a simple denoising network for effectively separating background and prospect in image denoising \cite{tian2020attention}. To improve performance, fusing hierarchical information enhances interaction of different layers to obtain more detailed information in image denoising \cite{zhang2018residual}. Besides, signal processing techniques embedded in a deep network can not only smooth images, but also can reduce complexity of a denoising network \cite{liu2018multi}. To reduce the computational cost of denoiser, a lightweight CNN with residual dense architecture is used to deal with color image demosaicking and denoising \cite{huang2018lightweight}. Mentioned denoising methods rely on improving network architectures to obtain better denoising effects for certain scenes, however, if a single feature from a CNN can effectively represent an image is key to a denoiser for complex scenes. 

In this paper, we present a cross Transformer denoising CNN (CTNet) containing a serial block (SB), a  parallel block (PB) and a residual block (RB) to obtain clean images for complex scenes. A SB depends on linearity and non-linearity components to deeply search structural information in image denoising. To mine more structural information based on different views, PB exploits three heterogeneous networks to implement interactions of multi-level features to broadly search information of more correlation between pixels for improving the adaptability of an obtained denoiser for complex scenes. To improve denoising performance, Transformer mechanisms are fused into the SB and PB to extract more salient features for filtering noise. Finally, a RB is exploited to predict clean images. Extended experiments report that the proposed CTNet is effective for image denoising in terms of evaluation of image denoising (i.e., quantitative and qualitative evaluation) and image quality assessment, and CTNet can process smartphone images. 

The proposed method can be summarized as the following contributions.  

1. Combining a serial and parallel architecture can search structural information in depth and breadth ways to improve denoising performance.

2. Three heterogeneous networks can achieve multi-level feature interactions to enhance relations of different networks for improving the adaptability of an obtained denoiser for complex scenes.   

3. Cross Transformer techniques in a CNN can extract more salient information to improve denoising performance, according to pixel relations. 

The second section provides related work. The third section presents the proposed method. The fourth section illustrates the analysis of the proposed method and experimental results. The final section reports the conclusion.

\section{Related work}
\subsection{Deep CNNs for image denoising}
Deep learning methods, i.e., CNNs rely on end-to-end architectures to obtain strong learning abilities that have been widely applied to image denoising \cite{tian2020deep}. They can be summarized as two categories: single CNN and multiple CNNs for image denoising.

Single CNN for image denoising: It is known that CNNs with single network architectures have obtained richer local features via translation invariance and smaller filters for image denoising \cite{tai2017memnet}.  Tai et al. relied on human thoughts to refer to a recursive unit and a gate unit to enhance memory ability of a deep network in image restoration \cite{tai2017memnet}. To reduce denoising time, a noise level mapping and noisy image patch are inputs of a deep CNN to train a denoiser for image denoising with different noise levels \cite{zhang2018ffdnet}. To improve denoising performance, dense connection operations and dilated convolutions are used to act on a CNN to address training difficulty and extract more useful information in image denoising \cite{chen2018deep}. Alternatively, Zhang et al. proposed a residual dense network architecture to obtain accurate hierarchical information to improve the quality of predicted images \cite{zhang2018residual}.  Quan et al. used a complex-valued transform to guide a CNN for obtaining more detailed information in image denoising \cite{quan2021image}. To obtain representative noise information, CNNs with multiple sub-networks are conducted for image denoising \cite{tian2020image}. 

Multiple CNNs for image denoising: To make obtained denoisers mine richer features for complex scenes, multiple CNNs are integrated as a unified architecture to effectively filter noise for image denoising. Tian et al. used renormalization batch techniques, dilated convolutions, and residual learning operations to achieve two heterogeneous network architectures to capture complementary structural information for obtaining clearer images \cite{tian2020image}. Taking relation between noise and texture into account, fusing multiscale operations, residual dense techniques and attention mechanisms into a dual CNN suppressed the speckle rather than damaging texture information for image denoising \cite{liu2021mrddanet}. To make a tradeoff between image denoising performance and efficiency, a sparse idea is exploited to guide dual CNN for obtaining an efficient denoising method for complex scenes \cite{tian2021designing}.  Besides, multiple CNNs are also extended in medical image denoising \cite{tojo2021medical}. According to mentioned illustrations, we can see that single network has faster execution denoising speed, and multiple CNNs are more stable for image denoising. Thus, we design a mixed denoising network based a serial and parallel architectures to extract more representative information for adaptively dealing with complex scenes in this paper.

\subsection{Attentive methods in image denoising}
Deep CNNs usually use backward propagations in gradient descent ways to obtain accurate information for image denoising \cite{tian2020deep}. To improve denoising efficiency, attention mechanism is exploited to quickly extract salient information for obtaining a denoiser \cite{tian2020attention}. It is divided into single head and multi-head attentions in image denoising. In terms of single head attention in image denoising, using front layer guides back layers in a network to agglomerate more useful information for efficiently removing noise \cite{tian2020deep}.  Tian et al. used a deep layer as current state to guide previous layer as previous state to separate foreground and background for suppressing noise \cite{tian2020attention}.  To enhance relations between different channels, an attention mechanism is used to act channels for removing redundant information in hyperspectral image denoising \cite{zhao2020adrn}.  Alternatively, spatial locations and channels are embedded into a frame via an attention mechanism to enhance effect of a certain convolutional layer for obtaining more useful structural information for image denoising \cite{wang2021channel}. Taking the effect of contextual information on noise into account, combining a pyramid technique, dilated convolution and residual learning operation into an attention method overcome drawback of contextual information for varying noise levels in a denoising model \cite{nikzad2021attention}. According to the multi-view idea, multi-head attention is presented in image denoising, which contains one network \cite{wang2022uformer} and multiple networks \cite{zhang2022multi}. Multi-head attention based one network uses different inputs to act on certain layer in a CNN to adjust effect of important information to improve efficiency of image denoising \cite{wang2022uformer}. A transformer is designed as a U-shaped attention mechanism to efficiently collect local context and adjust bias of multiple layers in the decoder for improving quality of predicted visual images \cite{wang2022uformer}. To quickly mine robust structural information, multi-head attention based multi networks exploited different inputs to act different networks in an attention manner to facilitate complementary salient information in image denoising \cite{zhang2022multi}.  Multiple rotated images are applied on a multi-head CNN in a multi-path attention to obtain image level features for suppressing noise \cite{zhang2022multi}. Besides, a dual attentive mechanism is deployed on two parallel sub-networks to enhance the correlation between spatial and spectral information on the 3D domain for hyperspectral image denoising \cite{shi2021hyperspectral}. According to the presentation above, we can see that the attention mechanism is a good choice for image denoising. However, it is noted that there is less interaction of different attention mechanisms to extract robust salient features in image denoising. Thus, we propose an interactive Transformer technique into a CNN to filter the noise in this paper.

\section{The proposed method}
\subsection{Network architecture}
To make a denoiser obtain more representative information for complex scenes, we propose a cross Transformer denoising CNN in image denoising (also referred to CTNet) as shown in Figure. 1. CTNet mainly uses interactions of different structural information obtained from different serial and parallel networks via attention manners in breadth and depth to extract salient features to better construct clean images. Its high denoising performance is implemented by a serial block (also referred to SB), a parallel block (also referred to PB) and a residual block (also referred to RB). A serial block uses linearity and non-linearity components to deeply search key information in image denoising. To obtain more complementary information based on different views, a parallel block is utilized to broadly search correlation information between pixels via interacting cross features obtained from three heterogeneous networks (i.e., SubNet1, SubNet2 and SubNet3) to improve adaptability of an obtained denoiser for complex scenes. To enhance robustness of the obtained denoiser, Transformer mechanisms are embedded into the SB and PB to extract more salient features for effectively filtering noise. Finally, CTNet exploits a residual block to covert obtained feature mapping to clean images. To intuitively understand the mentioned process, we conduct the following equation.
\begin{equation}
\begin{array}{ll}
{I_C} &= CTNet({I_N})\\
 &= {\rm{ }}RB{\rm{(}}PB{\rm{(}}SB{\rm{(}}{I_N}{\rm{)))}},
\end{array}
\end{equation}

\noindent where $I_N$ and $I_C$ are used to symbol a given noisy image and predicted clean image, respectively. $CTNet$ expresses a function of CTNet. $SB$, $PB$ and $RB$ are defined as functions of SB, PB and RB, respectively. More information of each key technique can be introduced in latter sub-sections.

\begin{figure*}[!htbp]
\centering
\subfloat{\includegraphics[width=6.5in]{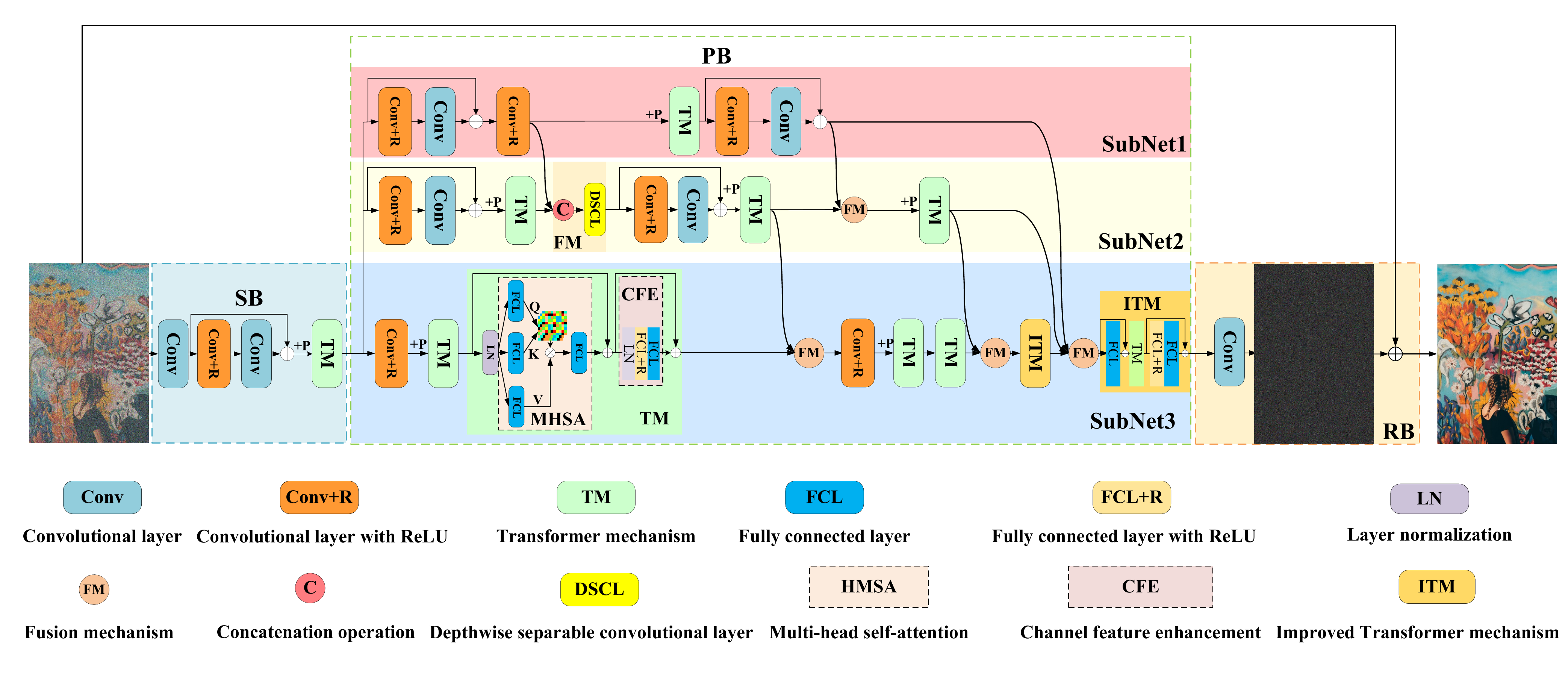}
  }
\caption{Network architecture of CTNet.}
 
\end{figure*}

\subsection{ Loss function}
To fairly verify denoising performance of our proposed CTNet, mean square error (MSE) \cite{allen1971mean} is chosen typical as image denoising methods, i.e., a denoising CNN (DnCNN) \cite{zhang2017beyond}, batch-renormalization denoising network (BRDNet) \cite{tian2020image}, fast and flexible denoising network (FFDNet) \cite{zhang2018ffdnet} is applied as an objective function to learn parameters. That is, training samples of paired $\{ I_C^i,I_N^i\} (1 \le i \le n)$ are utilized to act MSE to train a denoiser, where $I^i_C$ and $I^i_N$ are signified as the $ith$ clean image and corresponding noisy image in a training dataset. Also,  $n$ is the number of training samples. To simplify this process, we conduct the following equation. 

\begin{equation}
\begin{array}{ll}
l\left( \theta  \right){\rm{ }} = {\rm{ }}\frac{1}{{2n}}\sum\limits_{i = 1}^n {||CTNet(I_N^i) - I_C^i|{|^2}} 
\end{array}
\end{equation}
where $l$ is expressed as a loss function.  $\theta$ is implied as parameters of training a CTNet. Besides, Adam \cite{kingma2014adam} is acted on loss function to optimize parameters of training a CTNet in image denoising.

\subsection{Serial block}
SB is used to deeply search structural information in image denoising. Its good denoising performance relies on a serial architecture, which is mainly composed of three kinds, i.e., Conv, Conv+R and Transformer mechanism (TM). Conv is used as the first and third convolutional layers, where Conv denotes a convolutional layer. It can convert a given noisy image into linear features. Input and output channels of the first convolutional layer are 3 and 64 for color noisy images. Input and output channels of the first convolutional layers are 1 and 64 for gray noisy images. Also, Conv+R is the combination of linearity and non-linearity as the second convolutional layer to extract richer features, where Conv+R is defined as a combination of a convolutional layer and ReLU \cite{krizhevsky2017imagenet}. Conv is responsible for mining linear features. ReLU \cite{krizhevsky2017imagenet} is used as a piecewise function to transform linear information into non-linear features. Sizes of mentioned convolutional kernels are $3\times3$ . Taking the long-term dependency problem and robustness of obtained structural information into account, obtained linear and non-linear features of the second and third layers are fused via a residual learning operation as an input of transformer mechanism (TM). TM mainly depends on a Transformer \cite{chen2021pre} to mine relations between different patches to dynamically learn weights corresponding to different inputs for achieving an adaptive denoiser for different scenes. The mentioned illustrations can be formulated as follows.
\begin{equation}
\begin{array}{ll}
{O_{SB}} & = SB({I_N})\\
 & = TM(C(CR(C({I_N}))) + C({I_N}))\\
 & = TM({O_{IN\_TM}}), 
\end{array}
\end{equation}
where $C$ is a convolutional layer, $CR$ expresses a combination of a convolutional layer and ReLU,  $O_{IN\_TM}$is an input of the TM and $TM$ represents a transformer function, which is given more detailed information as follows. 

Taking denoising efficiency into account, a 4-layer TM \cite{chen2021pre} is implemented by an encoder rather than a combination of an encoder and a decoder in Figure.1. TM is composed of a 2-layer multi-head self-attention mechanism (MHSA) and a 2-layer channel feature enhancement (CFE). A 2-layer MHSA is used to extract global representations to enhance salient information for image denoising. That is, it uses a layer normalization (LN) \cite{ba2016layer} to unify distribution of obtained features from the input of TM as an input of three branches, where each branch consists of a 1-layer fully connected layer (FCL). Outputs of three branches are regarded as Q, K and V, respectively.  Also, we can integrate Q, K and V via an attention way to extract salient features, which acts a FCL to refine obtained features as output of MHSA. To enhance effect of shallow features, a residual operation is used to merge obtained features from the input of TM and output of MHSA as input of CFE. The mentioned process can be given in Eq. (4).

\begin{scriptsize}
\begin{equation}
\begin{array}{ll}
{O_{SB}} &= TM({O_{IN\_TM}}) \\
&= CFE(MHSA({O_{IN\_TM}})+ {O_{IN\_TM}})+O_{IN\_CFE}\\
&= {\rm{ }}CFE{\rm{(}}FCL{\rm{(softmax(}}\frac{{FCL{\rm{(}}LN({O_{IN\_TM}})) \times  FCL{{{\rm{(}}LN({O_{IN\_TM}}))}^T}}}{d}) \times FCL{\rm{(}}LN({O_{IN\_TM}}))) + {O_{IN\_TM}})+O_{IN\_CFE}\\
&= {\rm{ }}CFE{\rm{(}}FCL{\rm{(softmax(}}\frac{{{\rm{Q}} \times {K^T}}}{d}) \times V)+{O_{IN\_TM}})+O_{IN\_CFE}\\
&= {\rm{ }}CFE({O_{MHSA}}+ {O_{IN\_TM}})+O_{IN\_CFE}
\end{array}
\end{equation}
\end{scriptsize}

\noindent where $LN$ is a layer normalization operation, $d$ is a scaling factor, and $T$ stands for a transpose operation. $FCL$ denotes a fully connected layer and  $\times$ expresses a multiply operation as well as $ \otimes $ in Figure. 1. Also, $O_{MHSA}$ is an output of MHSA \cite{chen2021pre} and $O_{IN\_CFE}$ is 
 an input of CFE. CFE has two layers. The first layer of CFE is a stacked layer including a LN and FCL+R, where FCL+R is a combination of a FCL and a ReLU in Figure. 1. The second layer of CFE is a FCL. To inherit features of shallow layer, a residual learning operation is acted between an input of CFE and an output of the second layer. The mentioned illustrations can be presented in Eq. (5).

\begin{equation}
\begin{array}{ll}
{O_{SB}} &= CFE({O_{MHSA}} + {O_{IN\_TM}}) + O_{IN\_CFE}\\
 &= FCL(FCLR(LN({O_{MHSA}} + {O_{IN\_TM}}))) + O_{IN\_CFE}
\end{array}
\end{equation}
where $FCLR$ is FCL+R and $O_{SB}$ is an output of SB as an input of a parallel block. 

\subsection{Parallel block}
A 35-layer parallel block as well as PB is used to broadly search for more correlation information between pixels for improving the adaptability of a denoising model of CTNet for complex scenes. Its effectiveness is obtained by interactions of three heterogeneous networks (i.e., SubNet1, SubNet2 and SubNet3) in terms of multi-level features in Figure. 1. That is, three sub-networks interacted with each other to obtain complementary features from different views to enhance robustness of obtained denoising model. Specifically, a 9-layer SubNet1 consists of three types: Conv+R, Conv and TM. Conv+R is as the 1st, 3rd and 8th layers to extract non-linear information. Conv is used as the 2nd and 9th layers to extract linear information. TM is used to act between the 2nd and 3rd Conv+R to extract salient information. To enhance the expressive ability of a denoising network, the input of the 1st layer and output of the 2nd layer, and the input of the 8th layer and output of 9th layer are fused via residual learning operations, respectively. And all the convolutional kernel sizes are $3\times3$. Their input and output channels are 64, respectively. The mentioned illustrations can be presented as follows. 

\begin{equation}
\begin{array}{ll}
{O_{PB}} &= PB({O_{SB}})\\
 &= {\rm{ }}SubNet3{\rm{(}}{O_{SB}}{\rm{, }}SubNet1{\rm{(}}{O_{SB}}{\rm{)}}\Theta SubNet2{\rm{(}}{O_{SB}}{\rm{))}}\\
          &=  SubNet3\left( {{O_{SB}}{\rm{, }}{O_{SubNet1}},{\rm{ }}{O_{SubNet2}}} \right) 
\end{array}
\end{equation}
where $O_{PB}$ is output of PB, $SubNeti$ expresses SubNeti (i=1,2,3). $\Theta $ denotes an interaction operation. $O_{SubNet1}$ and $O_{SubNet2}$ stand for outputs of SubNet1 and SubNet2, respectively, where $O_{SubNet1}$ is obtained via Eq. (7). 

\begin{small}
\begin{equation}
    \begin{array}{ll}
    {O_{SubNet1}} &= SubNet1({O_{SB}})\\
    &= TM(CR(C(CR({O_{SB}})) + {O_{SB}}))+C(CR(TM(CR(C(CR({O_{SB}})) + {O_{SB}}))))\\
    &=  TM({O_{It}}) + C(CR(TM({O_{It}})))
    \end{array}
\end{equation}
\end{small}
where ${O_{It}} = CR(C(CR({O_{SB}}) + {O_{SB}}$ and $TM$  is introduced in Eq. (4). '+' denotes a residual learning, which is $\oplus$ in Figure. 1. +p denotes a position coding in Figure. 1. Also, $O_{It}$ and $O_{SubNet1}$ are used to interact SubNet2 and SubNet3 as follows. 

An 18-layer SubNet2 is implemented via four types, i.e., Conv+R, Conv, TM and depth-wise separable convolutional layer (DSCL).  Conv+R is used as the 1st and 8th layers to extract non-linear information. Conv constitutes the 2nd and 9th layers to obtain linear information, which is complementary to non-linear information. TM is acted as 3rd-6th, 10th-13th, 15th-18th layers to extract global representations for a robust denoiser, which has a detailed introduction in subsection Serial block. Depthwise separable convolutional layer as well as DSCL is the 7th and 14th layers in the SubNet2 to enhance relations of space in each channel. To address the long-term dependency of a deep network, two enhancement mechanisms are designed. The first mechanism uses a residual learning operation to fuse the input of the first layer and obtained features of the second layer as the input of the first TM in the SubNet2. The second mechanism utilizes a residual learning operation to merge obtained information of the 7th and 9th layers as the input of the second TM in the SubNet2. To enhance robustness of obtained features from obtained a denoiser, two interaction mechanisms are acted between SubNet1 and SubNet2. That is, the first interaction mechanism exploits obtained features of the third layer in the SubNet1 and the first TM in the SubNet2 as an input of the first DSCL. The second interaction mechanism utilizes outputs of the last layer in the SubNet1 and the second TM in the SubNet2 as an input of the second DSCL. The mentioned convolutional sizes of Conv+R and Conv are $3 \times 3$. Their input and output channels are 64, respectively. The illustrations above can be presented as the following Equation. 

\begin{footnotesize}
\begin{equation}
\begin{array}{ll}
     {{\rm{O}}_{SubNet2}} &= SubNet2({O_{SB}})\\
&= TM{\rm{(}}{O_{It2}}{\rm{)}} \\
&= TM(FM(TM(C(CR({O_{It3}})) + {O_{It3}}),{O_{SubNet1}})) \\
&= TM(FM(TM(C(CR(FM(TM(C(CR({O_{SB}}))+ {O_{SB}}),{O_{It}}))) + {O_{It3}}),{O_{SubNet1}}))
\end{array}
\end{equation}
\end{footnotesize}

\noindent where $O_{It2}$ is an output of the SubNet2 besides the last TM. $O_{It3}$ is the first interaction result between the SubNet1 and SubNet2, where ${O_{It3}} = FM(TM(C(CR({O_{SB}})) + {O_{SB}}),{O_{It}})$. Also, fusion mechanism (FM) is a combination of a concatenation operation and a DSCL. 

A 35-layer SubNet3 contains four types, i.e., Conv+R, TM, improved Transfomer mechanism (ITM) and DSCL. Conv+R is set as the 1st and 11th layers, which is also used to facilitate non-linear features. Also, all the convolutional kernel sizes are $3\times3$. Their input and output channels are 64, respectively.  TM is used as the 2th-9th and 12th-19th layers to obtain global representations to enhance salient information in image denoising, which is shown in Eq. (4). ITM \cite{chen2021pre} is utilized as the 21-27th and 29th-35th layers to eliminate interference information of previous different interactions, where ITM is stacked a FCL, TM and FCL+R and FCL. Also, to obtain more information, a residual learning operation is acted between input of the ITM and output of FCL, and outputs of the TM and FCL, respectively.  FM is acted as the 10th, 20th and 28th layers to interact with different networks to extract more robust features. The procedure can be given in Eq. (9). 

% \begin{small}
\begin{equation}
\begin{array}{ll}
{O_{PB}} &= SubNet3({O_{SB}},{O_{SubNet1}},{O_{SubNet2}})  \\
&= ITM(FM(ITM(FM(TM(TM(CR(FM(TM(TM(\\
&{}CR({O_{SB}}))),{O_{It2}})))),{O_{SubNet2}})),{O_{SubNet1}},{O_{SubNet2}}))
\end{array}
\end{equation}
% \end{small}

\noindent where $O_{PB}$ is the output of the PB to act the RB. 

\subsection {Residual block}

  A residual block as well as RB is composed of a 1-layer convolutional layer and a residual learning operation, which is used to construct a clean image by a residual learning operation to act an original noisy image and predicted image of the CTNet. Its convolutional kernel size is $3\times3$. Its input channel is 64 and output channel is decided by input images of the CTNet. When an input image is gray, the output channel is 1. When an input image is color, the output channel is 3. RB can be formulated as follows.

\begin{equation}
\begin{array}{ll}
{I_c} &= RB({O_{PB}})\\
&= {I_N} - C({O_{PB}})
\end{array}
\end{equation}
where – stands for a residual learning operation as well as $\oplus$ in Figure.1.  

\section{Experiments}
\subsection{Datasets}
Training datasets are divided into two parts: a synthetic noisy image training dataset and a real noisy image training dataset. According to Ref. \cite{liang2021swinir}, we choose Berkeley segmentation dataset (BSD) with 432 natural images \cite{martin2001database}, DIV2K with 800 natural images \cite{agustsson2017ntire}, Flickr2K with 2,650 natural images \cite{timofte2017ntire} and waterloo exploration database (WED) with 4,744 natural images \cite{ma2016waterloo} to form a synthetic noisy training dataset with 8,626 natural images. To enrich training datasets, each image is randomly cut into 108 image patches with a size of $48\times48$, which results in 931,608 image patches for training synthetic color and gray denoising models.  To train a real noisy image denoising model, we choose a real noisy image dataset with 100 natural images \cite{xu2018real} to conduct a real noisy image training dataset. Specially, each real noisy image is cut into 864,000 image patches with a size of $48 \times 48$. Additionally, to increase robustness of obtained denoising models, each image patch is randomly conducted by one way from eight ways from Ref. \cite{tian2020attention}.

Test datasets are composed of two parts: synthetic noisy image test datasets and real noisy image test datasets. Synthetic noisy image test datasets are chosen public datasets, i.e., BSD68 \cite{roth2005fields}, Set12 \cite{roth2005fields}, CBSD68 \cite{roth2005fields}, Kodak24 \cite{franzen1999kodak}, McMaster \cite{zhang2011color}, Urban100 \cite{huang2015single}, according to Ref. \cite{liang2021swinir}, to test gray and color Gaussian noisy image denoising models. Also, CC \cite{nam2016holistic} is used to test real noisy image denoiser, according to Ref. \cite{tian2020attention}. In addition, SIDD\cite{abdelhamed2018high} and real clinical dataset\cite{mccollough2017low} are selected to verify the effectiveness of the real scene.

\begin{table}[H]
\caption{PSNR (dB) results of some networks on BSD68 with $\sigma$ = 25.}
\centering
\begin{tabular}{|l|c|} 
\hline
\multicolumn{1}{|c|}{Methods}                                                                                                          & PSNR (dB)  \\ 
\hline
CTNet ~(Ours)                                                                                                                        & 29.464     \\ 
\hline
CTNet without TM in SB                                                                                                               & 29.442     \\ 
\hline
CTNet without TM and residual learning in SB \textbf{~}                                                                              & 29.436     \\ 
\hline
\begin{tabular}[c]{@{}l@{}}CTNet without TM and residual learning, and \\two ReLUs ~in SB\end{tabular}                                & 29.432     \\ 
\hline
CTNet with only
  three convolutional layers in SB                                                                                   & 29.428     \\ 
\hline
CTNet with only
  a convolutional layer in SB                                                                                        & 29.421     \\ 
\hline
Stacked a
  Conv in SB and a RB                                                                                                        & 26.182     \\ 
\hline
\begin{tabular}[c]{@{}l@{}}Stacked a Conv in SB and SubNet1 without two \\residual learning and a TM, and a RB\end{tabular}            & 28.878     \\ 
\hline
\begin{tabular}[c]{@{}l@{}}Stacked a Conv in SB and SubNet1 without a TM, \\and a RB\end{tabular}                                      & 28.884     \\ 
\hline
Stacked a
  Conv in SB and SubNet1                                                                                                     & 29.286     \\ 
\hline
\begin{tabular}[c]{@{}l@{}}Combination of a Conv in SB, SubNet1, SubNet2\\~without TM, residual learning, two FMs and a RB\end{tabular} & 29.311     \\ 
\hline
\begin{tabular}[c]{@{}l@{}}Combination of a Conv in SB, SubNet1, SubNet2\\~without TM, two FMs and a RB\end{tabular}                    & 29.322     \\ 
\hline
\begin{tabular}[c]{@{}l@{}}Combination of a Conv in SB, SubNet1, SubNet2 \\without two FMs, SubNet3, and a RB\end{tabular}              & 29.377     \\ 
\hline
\begin{tabular}[c]{@{}l@{}}Combination of a Conv in SB, SubNet1, SubNet2 \\without two FMs, and a RB\end{tabular}                                       & 29.338     \\ 
\hline
The combination of SB, SubNet3 and RB                                                                                                & 29.401     \\ 
\hline
\begin{tabular}[c]{@{}l@{}}The combination of SB, SubNet3 without TM and \\ITM, and RB\end{tabular}              & 29.284     \\ 
\hline
The combination of SB, and RB                                                                                                        & 29.219     \\ 
\hline
\begin{tabular}[c]{@{}l@{}}The combination of SB with SubNet3 without ITM \\in SubNet3, and RB\end{tabular}                            & 29.395     \\ 
\hline
CTNet
  without ITM                                                                                                                  & 29.430     \\ 
\hline
\begin{tabular}[c]{@{}l@{}}CTNet with a Conv in SB without the first two \\FMs in SubNet2 and SubNet3, and RB\end{tabular}          & 29.369     \\
\hline
\end{tabular}
\end{table}

\subsection{Experimental settings }
Parameters of training a CTNet denoiser are set as follows. Batch size is 8. The number of epochs is set to 33. Let initial learning rate be $2 \times 10^{-4} $, which varies from different epochs. That is, learning rate gets half when the epoch is the 15th, 22th, 24th, 26th, 28th, 30th and 31th epochs. $\beta1=0.9$ and $\beta2=0.99$. Other parameters can refer to Ref. \cite{tian2020attention}. Additionally, all the experiments are implemented by PyTorch 1.10.2 \cite{paszke2017pytorch} and Python 3.8.5, which run on Ubuntu 20.04 with AMD EPYC of 7502P/3.35GHz, 32-core CPU and RAM of 128G. To accelerate running speed, we use a GPU with a Nvidia GeForce GTX 3090, Nvidia CUDA 11.1 and cuDNN 8.04 to improve training efficiency of CTNet.

\subsection{Experimental analysis }
In this sub-section, we mainly analyze the rationality and validity of the proposed method composed of a serial block, a parallel block, and a residual block. It is known that deep networks depend on different layers to extract different hierarchical information for computer vision \cite{tian2020deep}. Robustness of obtained features from a single network architecture may be not guaranteed \cite{tian2020image}. Also, depth and breadth search methods from graph theory can use traversal ways to find useful information to the maximum extent \cite{rhee1994efficient}. Inspired by that, we use depth and breadth search ideas to guide a CNN for image denoising. A depth search method is used to direct a serial block to extract linear and non-linear features to improve robustness of the obtained denoiser. A breadth search method is utilized to guide a Parallel block to improve the adaptability of the obtained denoiser. Transformer mechanisms embedded in SB and PB can extract more salient information to effectively filter noise. Also, a RB is applied to construct clean images.  More design principles of these blocks can be presented as follows.  

\begin{figure*}[!htbp]
\centering
\subfloat{\includegraphics[width=1.0\linewidth]{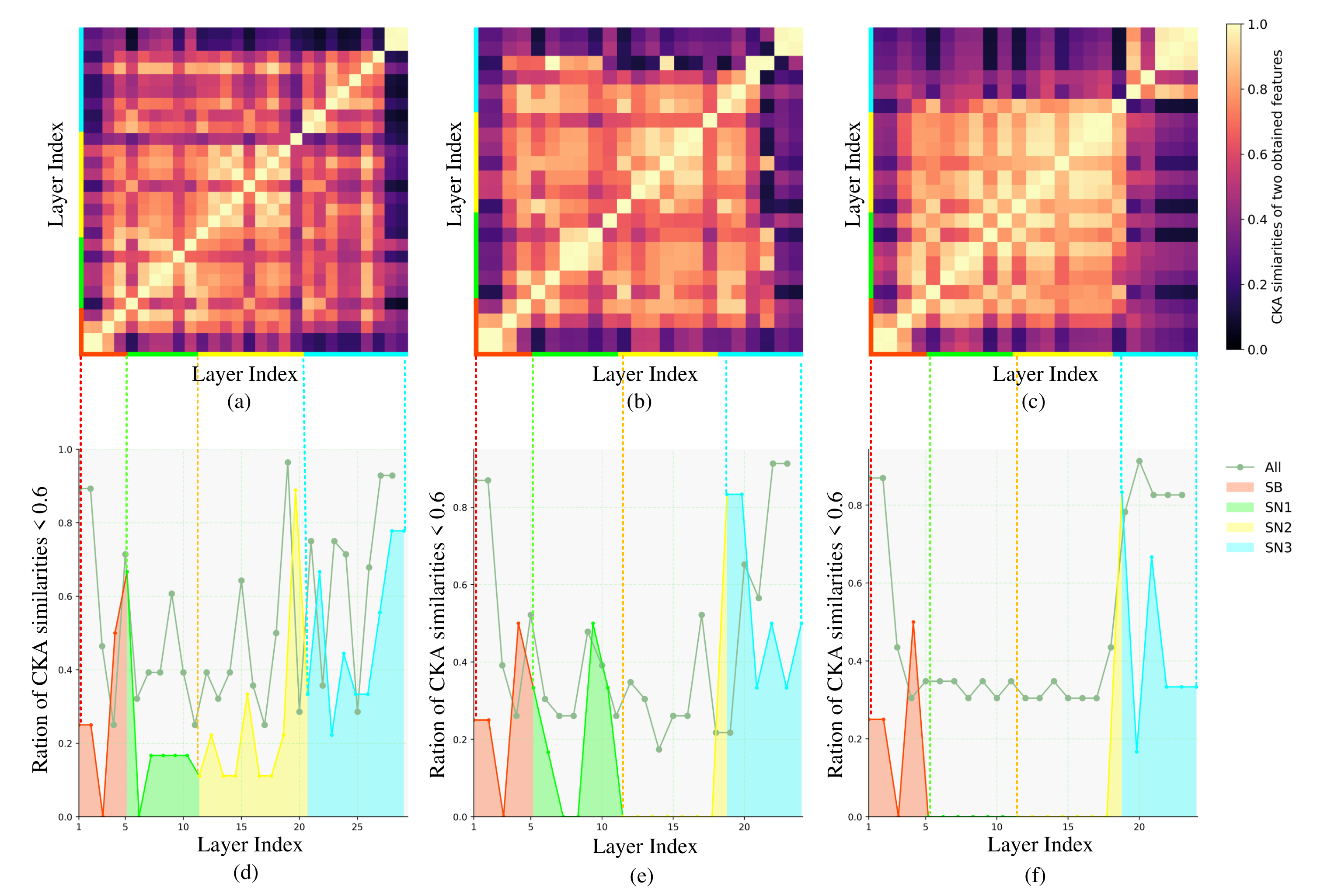}
  }
\caption{Sub-figures (a)-(c) illustrate CKA similarities of different methods in the CTNet, CTNet without the first two FMs in the SubNet2 and SubNet3, and CTNet with a serial architecture, respectively. Sub-figures (d)-(f) show ration of CKA similarities of different methods in the CTNet, CTNet without the first two FMs in the SubNet2 and SubNet3, and CTNet with a serial architecture, respectively.}
\end{figure*}

Serial block: It is known that deeper network architecture may cause gradient vanishing or explosion \cite{tian2020deep}. Also, depth search idea can deeply mine useful information to the maximum extent \cite{rhee1994efficient}. Thus, we use depth search idea to design a serial block to overcome mentioned problem and depth search path is chosen as follows. Firstly, we choose a stacked architecture (regarded as three stacked convolutional layers) to maintain performance superiority of a network, according to VGG \cite{simonyan2014very}, where these convolutional layers can extract structural information, where the first convolutional layer is used to convert a given noisy image to linear structural features. Its effectiveness is verified by ‘CTNet with only three convolutional layers in SB’ and ‘CTNet with only a convolutional layer in SB’ in Table 1.  Then, to make a trained model more robust, an activation function of ReLU was set behind the second layer to obtain more useful linear information. The design mainly considers denoising performance and efficiency. In terms of denoising performance, it is known that a heterogeneous architecture is useful to mine more accurate information in computer vision \cite{ren2019heterogeneous}. Thus, we only choose a ReLU behind the second convolutional layer rather than a combination of a convolutional layer and ReLU as the second and third layers. Its effectiveness is verified by the following steps. The first step uses ‘CTNet without TM and residual learning in SB’ and ‘CTNet with only  three convolutional layers in SB’ to show effectiveness of a ReLU acted in the SB in Table 1. The second uses ‘CTNet without TM and residual learning in SB’ and ‘CTNet without TM and residual learning, and two ReLUs in SB’ to show effectiveness of the mentioned heterogeneous architecture.  In terms of denoising efficiency, two ReLUs will reduce denoising time than that of a ReLU in a SB, which is verified by ‘CTNet without TM and residual learning in SB’ and ‘CTNet without TM and residual learning, and two ReLUs in SB’ in Table 2.  To improve deep search ability, a residual learning operation is used to fuse obtained features of the second and third layers in SB. Its superiority is shown via ‘CTNet without TM in SB’ and ‘CTNet without TM and residual learning in SB’ in Table 1. To overcome weakness of obtained structural information, a TM \cite{chen2021pre} is used to extract salient features in the end of the SB, according to similarity between different pixels. Its high performance is shown through the ‘CTNet’ and ‘CTNet without TM in SB’ in Table 1. Besides, ‘CTNet’ has higher peak signal-to-noise ratio (PSNR) \cite{korhonen2012peak} than that of ‘CTNet without TM in SB’, which shows good denoising performance of TM in SB. Also, the effectiveness of SB is verified by ‘CTNet’ and ‘CTNet with only a convolutional layer in SB' in Table 1. To improve robustness of obtained denoiser, a breadth search idea is used to guide a CNN to improving diversity of obtained information as follows.

\begin{table}
\caption{Denoising time of two networks on Urban100 with $\sigma$ = 15.}
\centering
\scalebox{0.9}{
\begin{tabular}{ccc} 
\toprule
Methods                                                         & $256\times256$ & $512\times512$  \\ 
\midrule
CTNet without TM and residual learning in SB               & 1.068s         & 4.454s          \\
CTNet without TM and residual learning, and \\ two ReLUs in SB & 1.073s         & 4.477s          \\
\bottomrule
\end{tabular}
}
\end{table}

\begin{table}
\caption{PSNR (dB) results of two networks on Urban100 with noise level of 15.}
\centering
\begin{tabular}{cc} 
\toprule
Methods                              & PSNR(dB)  \\ 
\hline
CTNet (Ours)                       & 33.722    \\
CTNet with a serial
  architecture & 33.590    \\
\bottomrule
\end{tabular}
\end{table}

Parallel block: It is known that deeper network architectures may cause gradient vanishing \cite{he2016deep}. Also, designing more heterogeneous architecture may improve image performance \cite{ren2019heterogeneous}. Thus, we use a breadth search idea rather than a depth search idea to implement a parallel block. That is, we use three sub-networks, i.e., SubNet1, SubNet2 and SubNet3 to conduct six interactions of multi-level features to broadly search important information for improving adaptability of an obtained denoiser for complex scenes. Specifically, the first network uses stacked Conv+R and Conv to extract diverse features, i.e., non-linear and linear features on the depth of a network to improve denoising performance.Its effectiveness is verified via ‘Stacked a Conv in SB and a RB’ and ‘Stacked a Conv in SB and SubNet1 without two residual learning and a TM, and a RB’ in Table 1. To make a trained model more robust, an activation function of ReLU was used to obtain more useful non-linear information. Its principle is explained in the SB in the fourth section. and the third section. In order to obtain salient features and eliminate redundant information, a TM is designed in SubNet1. Its effectiveness is verified by 'Stacked a Conv in SB and SubNet1' and 'Stacked a Conv in SB and SubNet1 without a TM, and a RB'. To improve the deep search ability, a residual learning operation is acted ends of each stacked Conv+R and Conv in SubNet1. Its effectiveness is verified via ‘Stacked a Conv in SB and SubNet1 without two residual learning and a TM, and a RB’ and ‘Stacked a Conv in SB and SubNet1 without a TM, and a RB’ in Table 1. To overcome limitations of inadequate features from single network, SubNet2 is designed as follows.

\begin{figure*}[!htbp]
\centering
\subfloat{\includegraphics[width=1.0\linewidth]{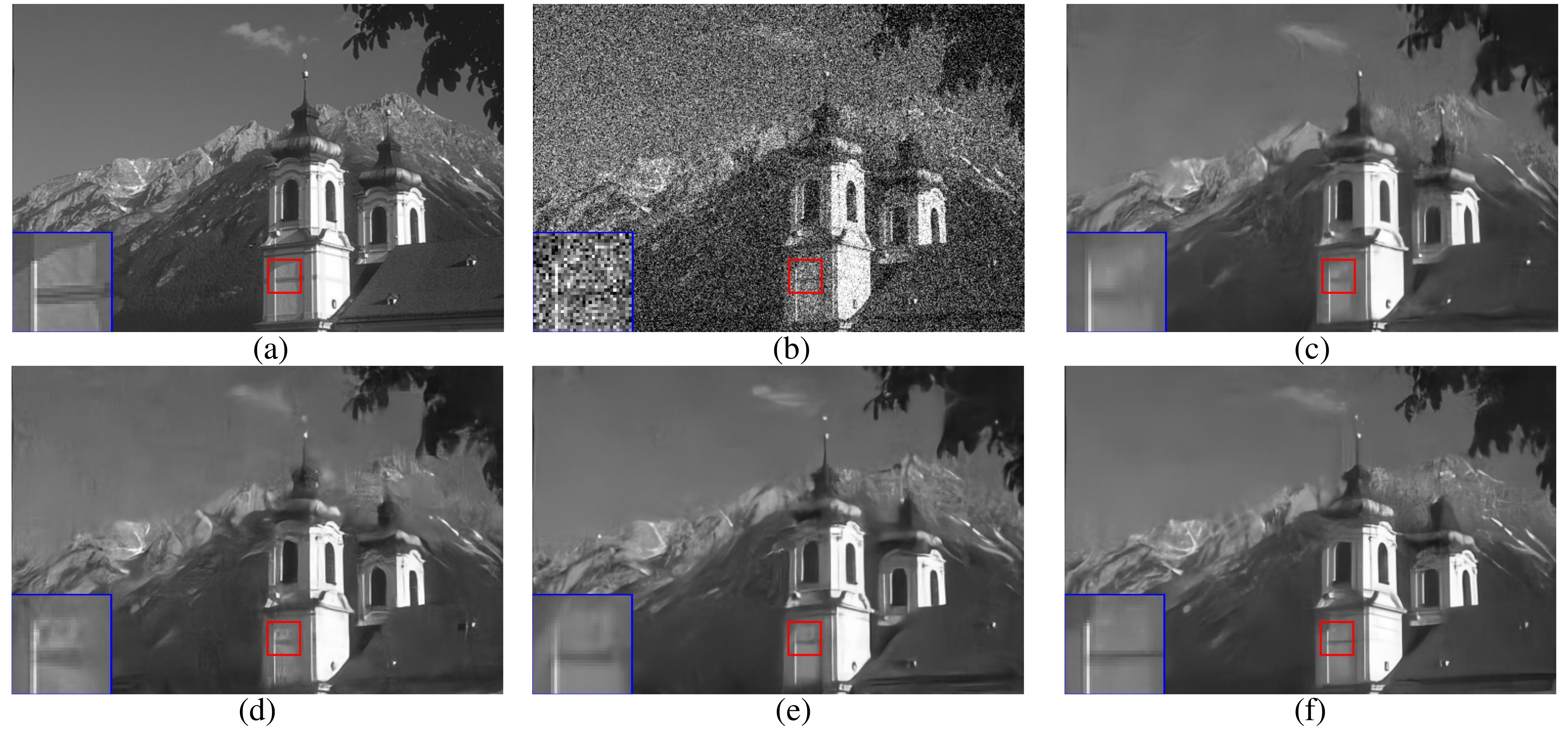}}
%\subfloat{\includegraphics[width=0.5\textwidth]{Fig4.pdf}}
\caption{Visual figures of different denoising methods on one image from BSD68 when $\sigma$ = 50. (a) Original image (b) Noisy image (c) ADNet/27.13 dB (d) DnCNN/27.92 dB (e) FFDNet/28.04 dB (f) CTNet (Ours)/28.16 dB.}
 
\end{figure*}

\begin{figure*}[!htbp]
\centering
\subfloat{\includegraphics[width=1.0\linewidth]{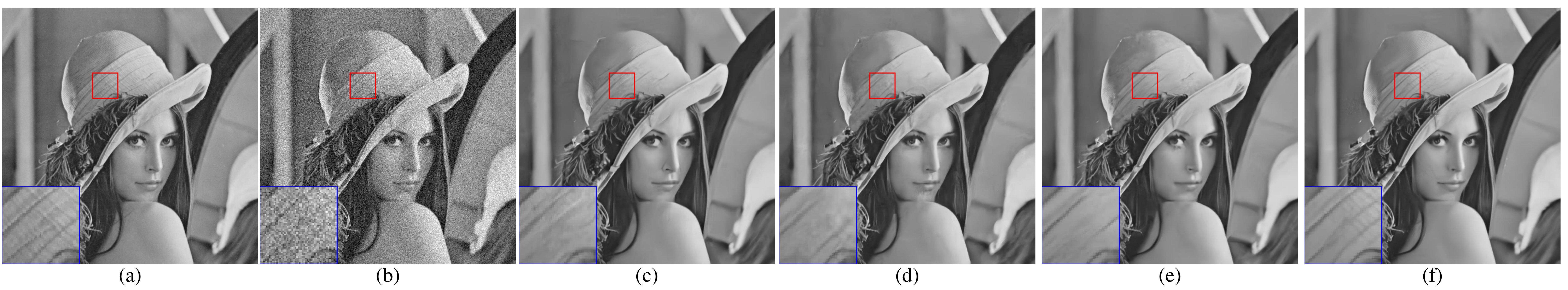}
  }
\caption{Visual figures of different denoising methods on one image from Set12 when $\sigma$ = 25. (a) Original image (b) Noisy image (c) ADNet/32.68 dB (d) DnCNN/32.44 dB (e) FFDNet/32.68 dB (f) CTNet(Ours)/32.87dB.}
 
\end{figure*}

\begin{table*}[!htbp]
\centering
\caption{Average PSNR (dB) of different methods on Set12 with noise levels of 15, 25 and 50.}
\scalebox{0.65}{
\begin{tabular}{cccccccccccccc} 
\toprule
Images                  & C.man                                                 & House                                & Peppers                              & Starfish                             & Monarch                              & Airplane                             & Parrot                                                & Lena                                                  & Barbara                              & Boat                                                  & Man                                                   & Couple                                                                   & Average \\ 
\hline
noise level             & \multicolumn{13}{c}{$\sigma=15$}\\ 
\hline
BM3D                    & 31.91                                                 & 34.93                                & 32.69                                & 31.14                                & 31.85                                & 31.07                                & 31.37                                                 & 34.26                                                 & 33.10                                & 32.13                                                 & 31.92                                                 & 32.10                                                                    & 32.37                                 \\
WNNM                    & 32.17                                                 & 35.13                                & 32.99                                & 31.82                                & 32.71                                & 31.39                                & 31.62                                                 & 34.27                                                 & 33.60                                & 32.27                                                 & 32.11                                                 & 32.17                                                                    & 32.70                                 \\
TNRD                    & 32.19                                                 & 34.53                                & 33.04                                & 31.75                                & 32.56                                & 31.46                                & 31.63                                                 & 34.24                                                 & 32.13                                & 32.14                                                 & 32.23                                                 & 32.11                                                                    & 32.50                                 \\
DnCNN                   & 32.61                                                 & 34.97                                & 33.30                                & 32.20                                & 33.09                                & 31.70                                & 31.83                                                 & 34.62                                                 & 32.64                                & 32.42                                                 & 32.46                                                 & 32.47                                                                    & 32.86                                 \\
FFDNet                  & 32.43                                                 & 35.07                                & 33.25                                & 31.99                                & 32.66                                & 31.57                                & 31.81                                                 & 34.62                                                 & 32.54                                & 32.38                                                 & 32.41                                                 & 32.46                                                                    & 32.77                                 \\
ECNDNet                 & 32.56                                                 & 34.97                                & 33.25                                & 32.17                                & 33.11                                & 31.70                                & 31.82                                                 & 34.52                                                 & 32.41                                & 32.37                                                 & 32.39                                                 & 32.39                                                                    & 32.81                                 \\
HDCNN                   & 32.51                                                 & 35.17                                & 33.22                                & 32.23                                & 33.20                                & 31.69                                & 31.86                                                 & 34.57                                                 & 32.60                                & 32.39                                                 & 32.36                                                 & 32.46                                                                    & 32.86                                 \\
BRDNet                  & \textcolor[rgb]{0,0.69,0.941}{32.80}                  & 35.27                                & 33.47                                & 32.24                                & 33.35                                & 31.85                                & 32.00                                                 & 34.75                                                 & 32.93                                & 32.55                                                 & 32.50                                                 & 32.62                                                                    & 33.03                                 \\
FOCNet                  & 32.71                                                 & 35.44                                & 33.41                                & 32.40                                & 33.29                                & 31.82                                & 31.98                                                 & 34.85                                                 & 33.09                                & 32.62                                                 & \textcolor[rgb]{0,0.69,0.941}{32.56}                  & 32.64                                                                    & 33.07                                 \\
CTNet (Ours)          & \textcolor{red}{32.82}                                & \textcolor{red}{35.86}               & \textcolor{red}{33.69}               & \textcolor{red}{32.65}               & \textcolor{red}{33.53}               & \textcolor{red}{32.07}               & \textcolor{red}{32.21}                                & \textcolor{red}{34.90}                                & \textcolor{red}{33.87}               & \textcolor{red}{32.75}                                & \textcolor{red}{32.61}                                & \textcolor{red}{32.77}                                                   & \textcolor{red}{33.31}                \\
CTNet-B (Ours)        & 32.66                                                 & \textcolor[rgb]{0,0.69,0.941}{35.68} & \textcolor[rgb]{0,0.69,0.941}{33.57} & \textcolor[rgb]{0,0.69,0.941}{32.52} & \textcolor[rgb]{0,0.69,0.941}{33.37} & \textcolor[rgb]{0,0.69,0.941}{31.96} & \textcolor[rgb]{0,0.69,0.941}{32.10}                  & \textcolor[rgb]{0,0.69,0.941}{34.87}                  & \textcolor[rgb]{0,0.69,0.941}{33.61} & \textcolor[rgb]{0,0.69,0.941}{32.66}                  & 32.55                                                 & \textcolor[rgb]{0,0.69,0.941}{32.70}                                     & \textcolor[rgb]{0,0.69,0.941}{33.19}  \\ 
\hline
noise level             & \multicolumn{13}{c}{$\sigma=25$}\\ 
\hline
BM3D                    & 29.45                                                 & 32.85                                & 30.16                                & 28.56                                & 29.25                                & 28.42                                & 28.93                                                 & 32.07                                                 & 30.71                                & 29.90                                                 & 29.61                                                 & 29.71                                                                    & 29.97                                 \\
WNNM                    & 29.64                                                 & 33.22                                & 30.42                                & 29.03                                & 29.84                                & 28.69                                & 29.15                                                 & 32.24                                                 & 31.24                                & 30.03                                                 & 29.76                                                 & 29.82                                                                    & 30.26                                 \\
TNRD                    & 29.72                                                 & 32.53                                & 30.57                                & 29.02                                & 29.85                                & 28.88                                & 29.18                                                 & 32.00                                                 & 29.41                                & 29.91                                                 & 29.87                                                 & 29.71                                                                    & 30.06                                 \\
DnCNN                   & 30.18                                                 & 33.06                                & 30.87                                & 29.41                                & 30.28                                & 29.13                                & 29.43                                                 & 32.44                                                 & 30.00                                & 30.21                                                 & 30.10                                                 & 30.12                                                                    & 30.43                                 \\
FFDNet                  & 30.10                                                 & 33.28                                & 30.93                                & 29.32                                & 30.08                                & 29.04                                & 29.44                                                 & 32.57                                                 & 30.01                                & 30.25                                                 & 30.11                                                 & 30.20                                                                    & 30.44                                 \\
ECNDNet                 & 30.11                                                 & 33.08                                & 30.85                                & 29.43                                & 30.30                                & 29.07                                & 29.38                                                 & 32.38                                                 & 29.84                                & 30.14                                                 & 30.03                                                 & 30.03                                                                    & 30.39                                 \\
DudeNet                 & 30.23                                                 & 33.24                                & 30.98                                & 29.53                                & 30.44                                & 29.14                                & 29.48                                                 & 32.52                                                 & 30.15                                & 30.24                                                 & 30.08                                                 & 30.15                                                                    & 30.52                                 \\
HDCNN                   & 30.03                                                 & 33.28                                & 30.75                                & 29.42                                & 30.37                                & 29.11                                & 29.43                                                 & 32.53                                                 & 30.03                                & 30.23                                                 & 30.01                                                 & 30.14                                                                    & 30.44                                 \\
N\textsuperscript{3}Net & 30.08                                                 & 33.25                                & 30.90                                & 29.55                                & 30.45                                & 29.02                                & 29.45                                                 & 32.59                                                 & 30.22                                & 30.26                                                 & 30.12                                                 & 30.12                                                                    & 30.55                                 \\
BRDNet                  & \textcolor{red}{31.39}                                & 33.41                                & 31.04                                & 29.46                                & 30.50                                & 29.20                                & 29.55                                                 & 32.65                                                 & 30.34                                & 30.33                                                 & 30.14                                                 & 30.28                                                                    & 30.61                                 \\
FOCNet                  & 30.35                                                 & 33.63                                & 31.00                                & 29.75                                & 30.49                                & 29.26                                & 29.58                                                 & 32.83                                                 & 30.74                                & 30.46                                                 & 30.22                                                 & 30.40                                                                    & 30.73                                 \\
CTNet (Ours)          & \textcolor[rgb]{0,0.69,0.941}{30.40}\textcolor{red}{} & \textcolor{red}{33.86}               & \textcolor{red}{31.33}               & \textcolor{red}{30.03}               & \textcolor{red}{30.68}               & \textcolor{red}{29.50}               & \textcolor{red}{29.73}                                & \textcolor{red}{32.87}                                & \textcolor{red}{31.62}               & \textcolor{red}{30.54}                                & \textcolor{red}{30.27}                                & \textcolor{red}{30.49}                                                   & \textcolor{red}{30.94}                \\
CTNet-B(Ours)        & 30.29\textcolor[rgb]{0,0.69,0.941}{}                  & \textcolor[rgb]{0,0.69,0.941}{33.78} & \textcolor[rgb]{0,0.69,0.941}{31.27} & \textcolor[rgb]{0,0.69,0.941}{29.92} & \textcolor[rgb]{0,0.69,0.941}{30.59} & \textcolor[rgb]{0,0.69,0.941}{29.45} & \textcolor[rgb]{0,0.69,0.941}{29.63}                  & \textcolor[rgb]{0,0.69,0.941}{32.84}                  & \textcolor[rgb]{0,0.69,0.941}{31.46} & \textcolor[rgb]{0,0.69,0.941}{30.50}                  & \textcolor[rgb]{0,0.69,0.941}{30.24}                  & \textcolor[rgb]{0,0.69,0.941}{30.44}                                     & \textcolor[rgb]{0,0.69,0.941}{30.87}  \\ 
\hline
noise level             & \multicolumn{13}{c}{$\sigma=50$}\\ 
\hline
BM3D                    & 26.13                                                 & 29.69                                & 26.68                                & 25.04                                & 25.82                                & 25.10                                & 25.90                                                 & 29.05                                                 & 27.22                                & 26.78                                                 & 26.81                                                 & 26.46                                                                    & 26.72                                 \\
WNNM                    & 26.45                                                 & 30.33                                & 26.95                                & 25.44                                & 26.32                                & 25.42                                & 26.14                                                 & 29.25                                                 & 27.79                                & 26.97                                                 & 26.94                                                 & 26.64                                                                    & 27.05                                 \\
TNRD                    & 26.62                                                 & 29.48                                & 27.10                                & 25.42                                & 26.31                                & 25.59                                & 26.16                                                 & 28.93                                                 & 25.70                                & 26.94                                                 & 26.98                                                 & 26.50                                                                    & 26.81                                 \\
DnCNN                   & 27.03                                                 & 30.00                                & 27.32                                & 25.70                                & 26.78                                & 25.87                                & 26.48                                                 & 29.39                                                 & 26.22                                & 27.20                                                 & 27.24                                                 & 26.90                                                                    & 27.18                                 \\
FFDNet                  & 27.05                                                 & 30.37                                & 27.54                                & 25.75                                & 26.81                                & 25.89                                & 26.57                                                 & 29.66                                                 & 26.45                                & 27.33                                                 & 27.29                                                 & 27.08                                                                    & 27.32                                 \\
ECNDNet                 & 27.07                                                 & 30.12                                & 27.30                                & 25.72                                & 26.82                                & 25.79                                & 26.32                                                 & 29.29                                                 & 26.26                                & 27.16                                                 & 27.11                                                 & 26.84                                                                    & 27.15                                 \\
DudeNet                 & 27.22                                                 & 30.27                                & 27.51                                & 25.88                                & 26.93                                & 25.88                                & 26.50                                                 & 29.45                                                 & 26.49                                & 27.26                                                 & 27.19                                                 & 26.97                                                                    & 27.30                                 \\
N\textsuperscript{3}Net & 27.14                                                 & 30.50                                & 27.58                                & 26.00                                & 27.03                                & 25.75                                & 26.50                                                 & 29.76                                                 & 27.01                                & 27.32                                                 & 27.33                                                 & 27.04                                                                    & 27.43                                 \\
HDCNN                   & 27.20                                                 & 30.04                                & 27.47                                & 25.73                                & 26.89                                & 25.82                                & 26.29                                                 & 29.50                                                 & 26.14                                & 27.16                                                 & 27.23                                                 & 26.93                                                                    & 27.20                                 \\
BRDNet                  & \textcolor[rgb]{0,0.69,0.941}{27.44}                  & 30.53                                & 27.67                                & 25.77                                & 26.97                                & 25.93                                & 26.66                                                 & 29.73                                                 & 26.85                                & 27.38                                                 & 27.27                                                 & 27.17                                                                    & 27.45                                 \\
FOCNet                  & 27.36                                                 & 30.91                                & 27.57                                & 26.19                                & \textcolor[rgb]{0,0.69,0.941}{27.10} & 26.06                                & \textcolor{red}{26.75}                                & \textcolor{red}{29.98}                                & 27.60                                & \textcolor{red}{27.53}                                & \textcolor{red}{27.42}                                & \textcolor{red}{27.39}                                                   & 27.68                                 \\
CTNet (Ours)          & \textcolor{red}{27.47}                                & \textcolor{red}{30.98}               & \textcolor{red}{27.92}               & \textcolor{red}{26.45}               & \textcolor{red}{27.14}               & \textcolor{red}{26.28}               & \textcolor[rgb]{0,0.69,0.941}{26.70}\textcolor{red}{} & \textcolor[rgb]{0,0.69,0.941}{29.89}\textcolor{red}{} & \textcolor{red}{28.29}               & \textcolor[rgb]{0,0.69,0.941}{27.52}\textcolor{red}{} & \textcolor[rgb]{0,0.69,0.941}{27.41}\textcolor{red}{} & \textcolor[rgb]{0,0.69,0.941}{27.37}\textcolor{red}{}                    & \textcolor{red}{27.79}                \\
CTNet-B(Ours)        & 27.43\textcolor[rgb]{0,0.69,0.941}{}                  & \textcolor[rgb]{0,0.69,0.941}{30.92} & \textcolor[rgb]{0,0.69,0.941}{27.89} & \textcolor[rgb]{0,0.69,0.941}{26.24} & 27.09\textcolor[rgb]{0,0.69,0.941}{} & \textcolor[rgb]{0,0.69,0.941}{26.27} & 26.66\textcolor[rgb]{0,0.69,0.941}{}                  & 29.86\textcolor[rgb]{0,0.69,0.941}{}                  & \textcolor[rgb]{0,0.69,0.941}{28.14} & 27.49\textcolor[rgb]{0,0.69,0.941}{}                  & 27.38\textcolor[rgb]{0,0.69,0.941}{}                  & \textcolor[rgb]{0.133,0.165,0.208}{27.34}\textcolor[rgb]{0,0.69,0.941}{} & \textcolor[rgb]{0,0.69,0.941}{27.73}  \\
\bottomrule
\end{tabular}
}
\end{table*}

\begin{table}[H]
\centering
\caption{Average PSNR (dB) of different methods on BSD68 with different noise levels of 15, 25 and 50.}
\begin{tabular}{cccc} 
\toprule
Methods                 & $\sigma$=15                   & $\sigma$=25                                 & $\sigma$=50                                  \\ 
\hline
BM3D                    & 31.07                  & 28.57                                & 25.62                                 \\
WNNM                    & 31.37                  & 28.83                                & 25.87                                 \\
TNRD                    & 31.42                  & 28.92                                & 25.97                                 \\
DnCNN                   & 31.73                  & 29.23                                & 26.23                                 \\
IRCNN                   & 31.63                  & 29.15                                & 26.19                                 \\
FFDNet                  & 31.63                  & 29.19                                & 26.29                                 \\
ECNDNet                 & 31.71                  & 29.22                                & 26.23                                 \\
DudeNet                 & 31.78                  & 29.29                                & 26.31                                 \\
ADNet                   & 31.74                  & 29.25                                & 26.29                                 \\
N\textsuperscript{3}Net & 31.78                  & 29.30                                & 26.39                                 \\
BRDNet                  & 31.79                  & 29.29                                & 26.36                                 \\
RIDNet                  & 31.81                  & 29.34                                & 26.40                                 \\
FOCNet                  & 31.83                  & 29.38                                & 26.50                                 \\
GCDN                    & 31.83                  & 29.35                                & 26.38                                 \\
MWCNN                   & 31.86                  & 29.41                                & 26.53                                 \\
NLRN                    & 31.88                  & 29.41                                & 26.47                                 \\
CDNet                   & 31.87                  & 29.41                                & \textcolor[rgb]{0,0.69,0.941}{26.52}  \\
MDRN+                   & 31.86                  & 29.39                                & 26.44                                 \\
DeamNet                 & 31.91                  & \textcolor[rgb]{0,0.69,0.941}{29.44} & \textcolor{red}{26.54}                \\
COLA-E                  & \textcolor[rgb]{0,0.69,0.941}{31.92}                  & \textcolor{red}{29.46}               & \textcolor[rgb]{0,0.69,0.941}{26.52}  \\
CTNet (Ours)          & \textcolor{red}{31.94} & \textcolor{red}{29.46}               & 26.49                                 \\
CTNet-B(Ours)         & 31.85\textcolor{red}{} & 29.39\textcolor[rgb]{0,0.69,0.941}{} & 26.45                                 \\
\bottomrule
\end{tabular}
\end{table}

SubNet2 also chooses an enhanced residual sub-architecture composed of two stacked Conv+R and Conv the same as SubNet1 to extract more richer structural information for image denoising, where a residual learning operation is acted input and output of each stacked Conv+R and Conv.  To further enlarge mining features ability of designed network, we propose a two-phase enhancement operation to obtain more effective information, according to breadth search idea. The first-phase enhancement uses a fusion mechanism composed of a concatenation operation and depthwise separable convolutional layer to fuse obtained features of SubNet1 and SubNet2. Also, ‘Combination of a Conv in SB, SubNet1, SubNet2 without TM, residual learning, two FMs and a RB’ has obtained higher PSNR than that of ‘Stacked a Conv in SB and SubNet1’ in Table 1, which shows effectiveness of two stacked convolutions in SubNet2 for image denoising. Superiority of two enhanced residual architectures is shown via ‘Combination of a Conv in SB, SubNet1, SubNet2 without TM, two FMs and a RB’ and ‘Combination of a Conv in SB, SubNet1, SubNet2 without TM, two FMs and residual learning, and a RB’ in Table 1. To prevent limitation of obtained hierarchical features, three TM operations is set into the SubNet2 to obtain extra feature in terms of pixel relation. That is, the first TM is acted between two enhanced residual sub-architectures. Other TM are stacked in the end of SubNet2. Their excellent denoising performance is listed by ‘Combination of a Conv in SB, SubNet1, SubNet2 without two FMs, and a RB’ and ‘Combination of a Conv in SB, SubNet1, SubNet2 without TM, two FMs and a RB’ in Table 1. To enhance relation of different layers of different networks, the second enhancement operation uses two FMs to respectively act between outputs of the third layer in SubNet1 and the first TM in SubNet2 and the last layer in SubNet1 and the second TM in SubNet2 to improve denoising effect, which is very effective via ‘CTNet with only a convolutional layer in SB’ and ‘Combination of a Conv in SB, SubNet1, SubNet2 without two FMs, SubNet3, and a RB’. Also, ‘CTNet’  has obtained higher PSNR than that ‘The combination of SB, SubNet3 and RB’ in Table 1, which shows good denoising effect of the combination of SubNet1 and SubNet2.  To enlarge relationship of hierarchical structural information and pixel features of images, we design the SubNet3 to pursue higher denoising results as follows.

\begin{table}[!htbp]
\caption{Average PSNR (dB) of different methods on Urban100 with different noise levels of 15, 25 and 50.}
\centering
\begin{tabular}{cccc} 
\toprule
Methods                  & $\sigma$=15                                 & $\sigma$=25                                 & $\sigma$=50                                  \\ 
\hline
BM3D                    & 32.35                                & 29.7                                 & 25.95                                 \\
WNNM                    & 32.97                                & 30.39                                & 26.83                                 \\
DnCNN                   & 32.64                                & 29.95                                & 26.26                                 \\
IRCNN                   & 32.46                                & 29.80                                & 26.22                                 \\
FFDNet                  & 32.43                                & 29.92                                & 26.52                                 \\
ADNet                   & 32.87                                & 30.24                                & 26.64                                 \\
N\textsuperscript{3}Net & 33.08                                & 30.19                                & 26.82                                 \\
RIDNet                  & 33.11                                & 30.49                                & 26.73                                 \\
FOCNet                  & 33.15                                & 30.64                                & 27.40                                 \\
GCDN                    & \textcolor[rgb]{0,0.69,0.941}{33.47} & 30.95                                & 27.41                                 \\
MWCNN                   & 33.17                                & 30.66                                & 27.42                                 \\
NLRN                    & 33.45                                & 30.94                                & 27.49                                 \\
CDNet                   & 33.36                                & 30.68                                & 27.02                                 \\
DeamNet                 & 33.37                                & 30.85                                & 27.53                                 \\
CTNet (Ours)          & \textcolor{red}{33.72}               & \textcolor{red}{31.28}               & \textcolor{red}{27.80}                \\
CTNet-B(Ours)         & 33.43                                & \textcolor[rgb]{0,0.69,0.941}{31.07} & \textcolor[rgb]{0,0.69,0.941}{27.68}  \\
\bottomrule
\end{tabular}
\end{table}

\begin{table}[!htbp]
\centering
\caption{Average PSNR (dB) of different methods on CBSD68 with different noise levels of 15, 25, 35, 50 and 75.}
\begin{tabular}{cccccc} 
\toprule
Methods          & $\sigma$=15                                                  & $\sigma$=25                                                  & $\sigma$=35                                                  & $\sigma$=50                                 & $\sigma$=75                                  \\ 
\hline
CBM3D           & 33.5                                                  & 30.69                                                 & 28.89                                                 & 27.36                                & 25.74                                 \\
IRCNN           & 33.86                                                 & 31.16                                                 & 29.5                                                  & 27.86                                & /                                     \\
FFDNet          & 33.87                                                 & 31.21                                                 & 29.57                                                 & 27.96                                & 26.24                                 \\
DnCNN           & 33.9                                                  & 31.24                                                 & 29.65                                                 & 27.95                                & /                                     \\
DSNet           & 33.91                                                 & 31.28                                                 & /                                                     & 28.05                                & /                                     \\
ADNet           & 33.99                                                 & 31.31                                                 & 29.66                                                 & 28.04                                & 26.33                                 \\
IRPDNN          & /                                                     & 31.24                                                 & 29.64                                                 & 28.06                                & 26.39                                 \\
GradNet         & 34.07                                                 & 31.39                                                 & /                                                     & 28.12                                & /                                     \\
CTCNN           & 33.98                                                 & 31.37                                                 & /                                                     & 28.15                                & /                                     \\
DudeNet         & 34.01                                                 & 31.34                                                 & 29.71                                                 & 28.09                                & 26.40                                 \\
BRDNet          & 34.10                                                  & 31.43                                                 & 29.77                                                 & 28.16                                & \textcolor[rgb]{0,0.69,0.941}{26.43}  \\
NBNet           & 34.15                                                 & 31.54                                                 & /                                                     & 28.35                                & /                                     \\
IPT             & /                                                     & /                                                     & /                                                     & 28.39                                & /                                     \\
NHNet           & 34.20                                                 & 31.54                                                 & /                                                     & 28.31                                & /                                     \\
DualBDNet       & \textcolor[rgb]{0,0.69,0.941}{34.32}                  & /                                                     & /                                                     & \textcolor{red}{28.61}               & /                                     \\
CTNet(Ours)         & \textcolor{red}{34.36}                                & \textcolor{red}{31.70}                                & \textcolor{red}{30.06}                                & \textcolor[rgb]{0,0.69,0.941}{28.43} & \textcolor{red}{26.67}                \\
CTNet-B(Ours) & \textcolor[rgb]{0,0.69,0.941}{34.32}\textcolor{red}{} & \textcolor[rgb]{0,0.69,0.941}{31.68}\textcolor{red}{} & \textcolor[rgb]{0,0.69,0.941}{30.04}\textcolor{red}{} & 28.41\textcolor[rgb]{0,0.69,0.941}{} & /\textcolor{red}{}                    \\
\bottomrule
\end{tabular}
\end{table}

SubNet3 uses two stacked Conv+R to extract non-linear hierarchical structural information. Its effectiveness is verified via ‘The combination of SB, SubNet3 without TM and ITM, and RB' and ‘The combination of SB, and RB’ in Table 1. Taking pixel relationship of images into account, we fuse four TM into a SubNet3 to extract more salient features. Specifically, two TM are acted between two stacked Conv+R. Also, other two TM are set behind the second Conv+R in the SubNet3. Its effectiveness is shown via ‘The combination of SB with SubNet3 without ITM in SubNet3, and RB’ and ‘The combination of SB with SubNet3 without TM and ITM in SubNet3, and RB’. To further improve denoising performance, we use ITM composed of FCL, two stacked FCL, TM and residual learning to facilitate more different features, i.e., linear and non-linear structural information from different layers, and salient structural information from pixel relations of images. Also, two ITM mechanisms is set in the end of the SubNet3. Its effectiveness is verified via ‘CTNet’ and ‘CTNet without ITM’.  To overcome inadequate interaction of SubNet1 and SubNet2, we use two FMs to transversely interact SubNet2 and SubNet3 to improve the robustness of obtained denoising model, according to breadth search. Its effectiveness is verified via ‘Combination of a Conv in SB, SubNet1, SubNet2 without two FMs, SubNet3, and a RB’ and ‘CTNet with a Conv in SB without the first two FMs in SubNet2 and SubNet3, and RB’.  Also, the combination of SubNet1, SubNet2, and SubNet3 is more effective than combination of SubNet1 and SubNet2 via ‘Combination of a Conv in SB, SubNet1, SubNet2 without two FMs, and a RB’ and ‘CTNet with a Conv in SB without the first two FMs in SubNet2 and SubNet3, and RB’ in Table 1. Additionally,  we use qualitative method to show effect of each layer on other layers as shown in Figure.2. That is, we count ration of centered kernel alignment (CKA) similarities \cite{li2021efficient} between current layer and other layers less than 0.6 to show the effect of current layer for image denoising. Figures.2 (a) and (d) demonstrate similarity of each layer and other layers in the CTNet. Figure.2 (b) and (e) demonstrate similarities of each layer and other layer in the CTNet without the first two FMs in the SubNet2 and SubNet3. As shown in Figures.2 (a) and (b), we can see that color of Figure. 2 (a) is heavier than Figure. (b), which shows that FM is very effective in our designed network for image denoising. Alternatively, we can see that SubNet2 (also regarded to SN2) and SubNet3 (also regraded to SN3) in the Figure.2 (d) have obtained higher ration value of CKA similarities than that of Figure.2 (e), which shows multiple interactions are important for image denoising. Finally, we explain reasons of combination of designed depth and breadth search architecture rather than only depth search architecture as follows. Firstly, deeper network architecture may cause gradient vanishing. Secondly, single network architecture may limit diversity of obtained features. Thus, we choose combination of designed depth and breadth search architecture to facilitate more robust information for image denoising. Its effectiveness is verified via ‘CTNet’ and ‘CTNet with a serial architecture’ in Table 3, where ‘CTNet with a serial architecture’ is stacked SB, SubNet1, SubNet2, SubNet3 and RB. Alternatively, we can see that color of Figure. 2 (a) is heavier than Figure. 2 (c), which shows that combination of designed depth and breadth search architecture is more effective in our designed network for image denoising.  Also, we can see that SubNet2 and SubNet3 in the Figure. 2 (d) have obtained higher ration value of CKA similarities than that of Figure. 2 (f), which shows effectiveness of combination of designed depth and breadth search architecture for image denoising. Besides, RB is only used to construct clean images. In summary, our proposed CTNet is reasonable and competitive for image denoising.
\begin{table}
\centering
\caption{Average PSNR (dB) of different methods on Kodak24 with different noise levels of 15, 25, 35, 50 and 75.}
\begin{tabular}{cccccc} 
\toprule
Methods          & $\sigma$=15                                                  & $\sigma$=25                                                  & $\sigma$=35                                                  & $\sigma$=50                                                  & $\sigma$=75                                  \\ 
\hline
CBM3D           & 34.26                                                 & 31.67                                                 & 30.56                                                 & 28.46                                                 & 26.82                                 \\
IRCNN           & 34.69                                                 & 32.18                                                 & 28.81                                                 & 28.93                                                 & /                                     \\
FFDNet          & 34.63                                                 & 32.13                                                 & 30.56                                                 & 28.98                                                 & 27.25                                 \\
DnCNN           & 34.60                                                  & 32.14                                                 & 30.64                                                 & 28.95                                                 & /                                     \\
DSNet           & 34.63                                                 & 32.16                                                 & /                                                     & 29.05                                                 & /                                     \\
ADNet           & 34.76                                                 & 32.26                                                 & 30.44                                                 & 29.10                                                  & 27.4                                  \\
IRPDNN          & /                                                     & 32.34                                                 & 30.81                                                 & 29.25                                                 & /                                     \\
GradNet         & 34.85                                                 & 32.35                                                 & /                                                     & 29.23                                                 & /                                     \\
CTCNN           & 34.65                                                 & 32.34                                                 & /                                                     & 29.22                                                 & /                                     \\
NHNet           & 35.02                                                 & 32.54                                                 & /                                                     & 29.41                                                 & /                                     \\
DudeNet         & 34.81                                                 & 32.26                                                 & 30.69                                                 & 29.10                                                 & 27.39                                 \\
BRDNet          & 34.88                                                 & 32.41                                                 & 30.80                                                 & 29.22                                                 & \textcolor[rgb]{0,0.69,0.941}{27.72}  \\
IPT             & /                                                     & /                                                     & /                                                     & 29.64                                                 & \textcolor[rgb]{0,0.69,0.941}{/}      \\
CTNet (Ours)  & \textcolor{red}{35.28}                                & \textcolor{red}{32.82}                                & \textcolor{red}{31.26}                                & \textcolor{red}{29.67}                                & \textcolor{red}{27.91}                \\
CTNet-B(Ours) & \textcolor[rgb]{0,0.69,0.941}{35.24}\textcolor{red}{} & \textcolor[rgb]{0,0.69,0.941}{32.79}\textcolor{red}{} & \textcolor[rgb]{0,0.69,0.941}{31.24}\textcolor{red}{} & \textcolor[rgb]{0,0.69,0.941}{29.65}\textcolor{red}{} & /\textcolor{red}{}                    \\
\bottomrule
\end{tabular}
\end{table}

\begin{table}
\caption{Average PSNR (dB) of different methods on McMaster with different noise levels of 15, 25, 35, 50 and 75.}
\centering
\begin{tabular}{cccccc} 
\toprule
Methods          & $\sigma$=15                                                  & $\sigma$=25                                                  & $\sigma$=35                                                  & $\sigma$=50                                                  & $\sigma$=75                                  \\ 
\hline
CBM3D           & 34.03                                                 & 31.63                                                 & 29.92                                                 & 28.48                                                 & 26.79                                 \\
IRCNN           & 34.58                                                 & 32.18                                                 & 30.59                                                 & 28.91                                                 & /                                     \\
FFDNet          & 34.66                                                 & 32.35                                                 & 30.76                                                 & 29.18                                                 & 27.29                                 \\
DnCNN           & 33.45                                                 & 31.52                                                 & 30.91                                                 & 28.62                                                 & /                                     \\
DSNet           & 34.67                                                 & 32.40                                                 & /                                                     & 29.28                                                 & /                                     \\
ADNet           & 34.93                                                 & 32.56                                                 & 31.00                                                 & 29.36                                                 & 27.53                                 \\
IRPDNN          & /                                                     & 32.33                                                 & 30.90                                                 & 29.33                                                 & \textcolor[rgb]{0,0.69,0.941}{27.59}  \\
GradNet         & 34.81                                                 & 32.45                                                 & /                                                     & 29.39                                                 & ~                                     \\
BRDNet          & 35.08                                                 & 32.75                                                 & 31.15                                                 & 29.52                                                 & 27.27                                 \\
IPT             & /                                                     & /                                                     & /                                                     & 29.98                                                 & /                                     \\
CTNet (Ours)  & \textcolor{red}{35.54}                                & \textcolor{red}{33.21}                                & \textcolor{red}{31.67}                                & \textcolor{red}{30.02}                                & \textcolor{red}{28.15}                \\
CTNet-B(Ours) & \textcolor[rgb]{0,0.69,0.941}{35.46}\textcolor{red}{} & \textcolor[rgb]{0,0.69,0.941}{33.17}\textcolor{red}{} & \textcolor[rgb]{0,0.69,0.941}{31.64}\textcolor{red}{} & \textcolor[rgb]{0,0.69,0.941}{30.00}\textcolor{red}{} & /\textcolor{red}{}                    \\
\bottomrule
\end{tabular}
\end{table}

\subsection{Experimental results}

To fully test denoising effect of the proposed CTNet, we apply quantitative and qualitative metrics to evaluate it. Quantitative evaluation  mainly uses some state-of-the-arts, i.e., BM3D\cite{dabov2007image}, WNNM \cite{gu2014weighted}, trainable nonlinear reaction diffusion (TNRD)\cite{chen2016trainable}, DnCNN\cite{zhang2017beyond}, image restoration CNN (IRCNN) \cite{zhang2017learning}, FFDNet\cite{zhang2018ffdnet}, enhanced convolutional neural denoising network (ECNDNet)\cite{tian2019enhanced}, Dual denoising Network (DudeNet) \cite{tian2021designing}, attention-guided denoising convolutional neural network (ADNet)\cite{tian2020attention}, neural nearest neighbors networks (N$^3$Net)\cite{plotz2018neural}, BRDNet\cite{tian2020image}, single-stage blind real image denoising network (RIDNet)\cite{anwar2019real}, fractional optimal control network (FOCNet)\cite{jia2019focnet}, graph convolution image denoising network (GCDN)\cite{valsesia2020deep}, multi-level wavelet CNN (MWCNN)\cite{liu2018multi}, non-local recurrent network (NLRN)\cite{liu2018non}, a hybrid denoising CNN (HDCNN)\cite{zheng2022hybrid}, compact denoising network (CDNet)\cite{ko2022blind}, multiple degradation and reconsturctural etwork with self-ensemble mechanism (MDRN+)\cite{li2022multiple}, a network with dual element-wise attention mechanism (DeamNet)\cite{ren2021adaptive}, collaborative attention network (COLA-E)\cite{mou2021cola}, dynamic attentive graph learning model (DAGL)\cite{mou2021dynamic}, Dilated Residual Networks with Symmetric Skip Connection (DSNet)\cite{peng2019dilated}, Identifying Recurring Patterns with Deep Neural Networks (IRPDNN)\cite{xia2020identifying}, image denoising network with gradient (GradNet)\cite{liu2020gradnet}, Countourlet Transform based convolutional network model (CTCNN)\cite{lyu2020nonsubsampled}, noise basis learning network (NBNet)\cite{cheng2021nbnet}, non‐local hierarchical network (NHNet)\cite{zhang2022nhnet}, dynamic dual learning network for blind image denoising (DualBDNet)\cite{du2020blind}, targeted image denoising (TID)\cite{luo2015adaptive} and generative adversarial capsule network (DeCapsGAN)\cite{lyu2021decapsgan}, image processing transformer (IPT)\cite{chen2021pre} on noisy images (i.e., synthetic and real noisy images) for image denoising. Specifically, synthetic noisy images contain gray and color synthetic noisy images. 

\begin{table*}
\caption{PSNR (dB) results of different methods on real noisy images.}
\scalebox{0.8}{
\centering

\begin{tabular}{ccccccccc} 
\toprule
Setting                                & \begin{tabular}[c]{@{}c@{}}CBM3D ~\end{tabular} & TID   & DnCNN & ADNet                                & DudeNet & DeCapsGAN                            & BRDNet                               & CTNet(ours)                                          \\ 
\hline
\multirow{3}{*}{Canon 5D ISO = 3200}   & \textcolor{red}{39.76}                            & 37.22 & 37.26 & 35.96                                & 36.66   & 35.74                                & 37.63                                & \textcolor[rgb]{0,0.69,0.941}{39.27}\textcolor{red}{}  \\
                                       & 36.40                                             & 34.54 & 34.13 & 36.11                                & 36.70   & 37.02                                & \textcolor{red}{37.28} & \textcolor[rgb]{0,0.69,0.941}{37.08}                                 \\
                                       & 36.37                                             & 34.25 & 34.09 & 34.49                                & 35.03   & 36.47                                & \textcolor{red}{37.75}               & \textcolor[rgb]{0,0.69,0.941}{36.57}\textcolor{red}{}  \\ 
\hline
\multirow{3}{*}{Nikon D600 ISO = 3200} & 34.18                                             & 32.99 & 33.62 & 33.94                                & 33.72   & \textcolor{red}{35.71}               & 34.55                                & \textcolor[rgb]{0,0.69,0.941}{35.46}\textcolor{red}{}  \\
                                       & 35.07                                             & 34.20 & 34.48 & 34.33                                & 34.70   & 35.83                                & \textcolor[rgb]{0,0.69,0.941}{35.99} & \textcolor{red}{36.75}                                 \\
                                       & 37.13                                             & 35.58 & 35.41 & \textcolor[rgb]{0,0.69,0.941}{38.87} & 37.98   & 36.93                                & 38.62                                & \textcolor{red}{41.04}                                 \\ 
\hline
\multirow{3}{*}{Nikon D800 ISO = 1600} & 36.81                                             & 34.49 & 37.95 & 37.61                                & 38.10   & 38.41                                & \textcolor[rgb]{0,0.69,0.941}{39.22} & \textcolor{red}{40.12}                                 \\
                                       & 37.76                                             & 35.19 & 36.08 & 38.24                                & 39.15   & 39.14                                & \textcolor[rgb]{0,0.69,0.941}{39.67} & \textcolor{red}{41.59}                                 \\
                                       & 37.51                                             & 35.26 & 35.48 & 36.89                                & 36.14   & 37.19                                & \textcolor[rgb]{0,0.69,0.941}{39.04} & \textcolor{red}{39.95}                                 \\ 
\hline
\multirow{3}{*}{Nikon D800 ISO = 3200} & 35.05                                             & 33.70 & 34.08 & 37.20                                & 36.93   & 37.68                                & \textcolor[rgb]{0,0.69,0.941}{38.28} & \textcolor{red}{39.88}                                 \\
                                       & 34.07                                             & 31.04 & 33.70 & 35.67                                & 35.80   & 36.85                                & \textcolor[rgb]{0,0.69,0.941}{37.18} & \textcolor{red}{38.45}                                 \\
                                       & 34.42                                             & 33.07 & 33.31 & 38.09                                & 37.49   & 36.85                                & \textcolor[rgb]{0,0.69,0.941}{38.85} & \textcolor{red}{40.44}                                 \\ 
\hline
\multirow{3}{*}{Nikon D800 ISO = 6400} & 31.13                                             & 29.40 & 29.83 & 32.24                                & 31.94   & \textcolor[rgb]{0,0.69,0.941}{33.32} & 32.75                                & \textcolor{red}{34.01}                                 \\
                                       & 31.22                                             & 29.86 & 30.55 & 32.59                                & 32.51   & 31.81                                & \textcolor[rgb]{0,0.69,0.941}{33.24} & \textcolor{red}{34.26}                                 \\
                                       & 30.97                                             & 29.21 & 30.09 & 33.14                                & 32.91   & \textcolor{red}{33.67} & 32.89                                & \textcolor[rgb]{0,0.69,0.941}{33.48}                                 \\ 
\hline
Average                                & 35.19                                             & 33.63 & 33.86 & 35.69                                & 35.72   & 36.18                                & \textcolor[rgb]{0,0.69,0.941}{36.73} & \textcolor{red}{37.89}                                 \\
\bottomrule
\end{tabular}
}
\end{table*}

For gray synthetic noisy image denoising, we choose Set12 \cite{roth2005fields}, BSD68 \cite{roth2005fields} and Urban100 \cite{huang2015single} to evaluate denoising performance of our CTNet as shown in Tables 4-6. As presented in mentioned three Tables, our method nearly has obtained the best denoising performance in terms of PSNR for noise levels of 15, 25 and 50. For example, our CTNet has exceeded PSNR of 0.12dB than that of the second method for noise level of 15, where the second method is our CTNet for blind denoising (CTNet-B) and CTNet-B is trained via varying noise levels from 0 to 55. That illustrates our CTNet is very suitable for image blind denoising.  Our CTNet has obtained improvement of 0.51dB than that baseline of denoising method, i.e., DnCNN for noise level of 25 in Table 4. Our CTNet has obtained improvement of 0.26dB than that baseline of denoising method, i.e., DnCNN for image denoising in Table 5.  Also, our CTNet has obtained improvement of 0.25dB than that of the second GCDN for noise level of 15 in Table 6. It also has improvement of 1.08dB than that of baseline of denoising method, i.e., DnCNN for noise level of 15 in Table 6. These show that our method is very effective for gray noisy image denoising. For color synthetic noisy image denoising, we conducted some experiments on CBSD68 \cite{roth2005fields}, Kodak24 \cite{franzen1999kodak} and McMaster \cite{zhang2011color} for noise levels of 15, 25, 35, 50 and 75. As listed in Tables 7-9, we can see that our method nearly obtained the best denoising results for all the noise levels. For example, our CTNet has exceeded 0.19dB than that of the second method on noise level of 75 for color synthetic noisy image denoising in Table 8. And it has improvement of 0.62dB than that of baseline of denoising method, i.e., DnCNN on noise level of 35 for color synthetic noisy image denoising in Table 8. Also, our method has obtained better denoising results than that of popular denoising method of IPT based Transformer as shown in Tables 7-9 and our method only takes parameters and Flops of 44\% of the IPT in Table 11. That shows our method is useful for color synthetic noisy image denoising.

\begin{table}
\caption{Complexities of two denoising networks.}
\centering
\begin{tabular}{ccc} 
\toprule
Methods & Parameters & Flops   \\ 
\hline
IPT     & 109.97M    & 15.37G  \\
CTNet & 49.03M     & 6.91G   \\
\bottomrule
\end{tabular}
\end{table}

\begin{table}
\caption{PSNR (dB) results of different methods for real noisy images on SIDD\cite{abdelhamed2018high} and Mayo datasets\cite{mccollough2017low}.}
\centering
\begin{tblr}{
  row{1} = {c},
  row{2} = {c},
  cell{1}{1} = {r=2}{},
  cell{1}{2} = {c=2}{},
  cell{3}{2} = {c},
  cell{3}{3} = {c},
  cell{4}{2} = {c},
  cell{4}{3} = {c},
  cell{5}{2} = {c},
  cell{5}{3} = {c},
  hline{1,6} = {-}{0.08em},
  hline{2} = {2-3}{},
  hline{3} = {-}{},
}
Methods     & PSNR &    \\
            & SIDD & Mayo \\
DnCNN       & ~\textcolor[rgb]{0,0.69,0.941}{26.21}   & ~21.86 \\
IPT          & ~26.08   & ~\textcolor[rgb]{0,0.69,0.941}{23.30} \\
CTNet(Ours) & ~\textcolor{red}{28.05}   & ~\textcolor{red}{24.13} 
\end{tblr}
\end{table}

For real noisy image denoising, we choose some popular denoising methods, i.e., CBM3D, TID, DnCNN, ADNet, DudeNet, DeCapsGAN and BRDNet on CC\cite{nam2016holistic} with different ISO to test denoising performance of the proposed CTNet. As shown in Table 10, our method nearly has obtained the best PSNR for most of ISO. For example, our method has exceeded 1.16dB than that of the second denoising method, i.e., BRDNet. Also, our method has an improvement of 3.39dB than that of baseline of denoising method, i.e. DnCNN for Nikon D800 ISO = 6400. That shows that our method is effective for real noisy image denoising with different ISO. As shown in Table 12, our method outperforms other denoising methods on smartphone images and CT images and has an improvement of 1.84 dB and 2.27dB than DnCNN, respectively. That shows our method is very effective for real noisy image denoising. Besides, red and blue lines are the best and second denoising effects from Table 4 to Table 12 besides Table 11. According to mentioned illustrations, it is known that our method is very competitive in terms of quantitative evaluation. 

\begin{figure*}[!htbp]
\centering
\subfloat{\includegraphics[width=1.0\linewidth]{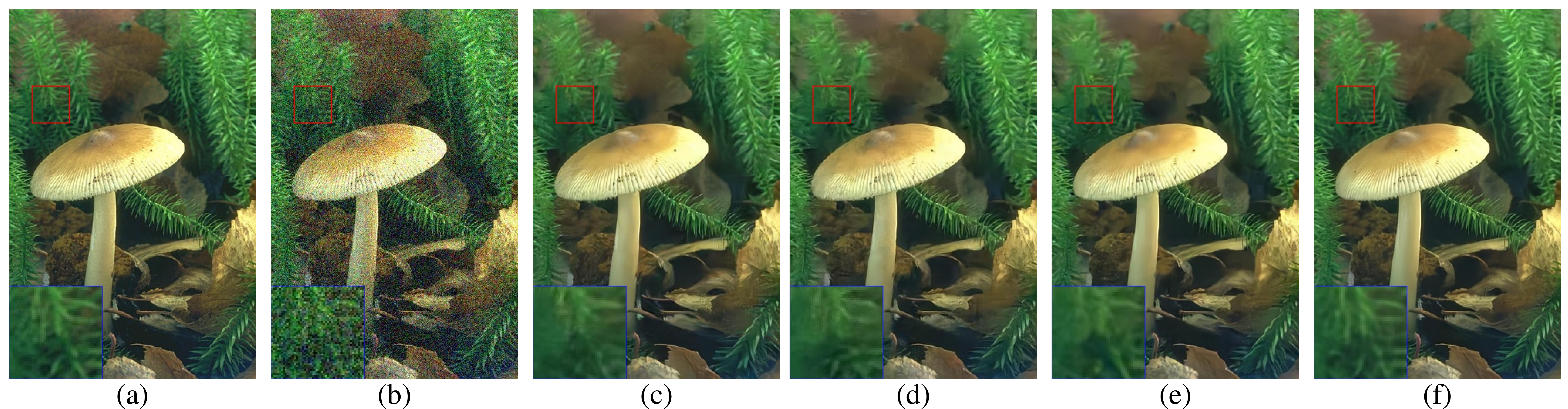}}
% \subfloat{\includegraphics[width=0.5\textwidth]{Fig5.pdf}}
\caption{Visual figures of different denoising methods on one image from CBSD68 when $\sigma$ = 35. (a) Original image, (b) Noisy image, (c) ADNet/28.84 dB, (d) DnCNN/29.33 dB, (e) FFDNet/29.30 dB and (f) CTNet (Ours)/29.90 dB.}
 
\end{figure*}

\begin{figure*}[!htbp]
\centering
\subfloat{\includegraphics[width=1.0\linewidth]{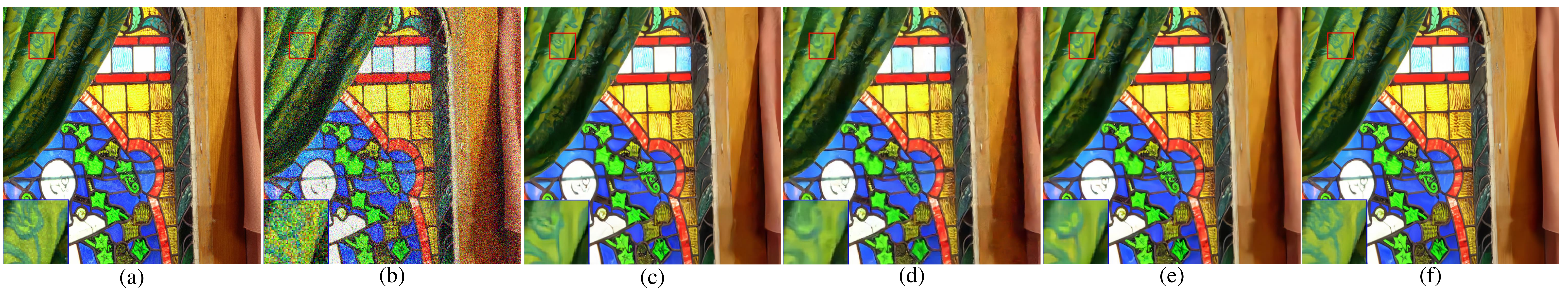}}
\caption{Visual figures of different denoising methods on one image from McMaster when $\sigma$ = 50. (a) Original image, (b) Noisy image, (c) ADNet/23.70 dB, (d) DnCNN/24.07 dB, (e) FFDNet/25.56 dB and (f) CTNet (Ours)/26.39 dB.} 
\end{figure*}

For qualitative evaluation, we use four popular denoising methods, i.e., ADNet, DnCNN, FFDNet as comparative methods on Set12, BSD68, CBSD68 and McMaster to conduct visual figures for testing denoising effects of the proposed CTNet. That is, we choose an area of denoising images from different methods as an observation area, the observation area is clearer, and its corresponding method is more effective for image denoising. As shown in Figures. 3 and 4, we can see that the proposed CTNet has obtained clearer area than that of baseline of denoising methods, i.e., ADNet, DnCNN and FFDNet. That shows our method is very effective for gray noisy image denoising. As illustrated in Figures. 5 and 6, we can see that our CTNet has obtained more detailed information than other denoising methods. That reports our method is very suitable to color noisy image denoising. According to the mentioned illustrations, it is known that our CTNet is very competitive for image denoising in terms of quantitative and qualitative evaluations. 

\section{Conclusion}
In this paper, we propose a cross Transformer denoising CNN composed of a serial block, a parallel block and a residual block for image denoising. A SB can guide an enhanced residual architecture via depth search ideas to facilitate structural information for image denoising. To avoid loss of key information, PB designs three heterogeneous networks to implement multiple interactions of different features to extract richer detailed information in image denoising, according to broadly search idea. Also, to improve denoising performance, Transformer mechanisms are embedded into the SB and PB to extract complementary salient information for effectively removing noise. Finally, a RB is applied to acquire clean images. The proposed method is very suitable to image denoising for complex scenes, according to a lot of experimental analysis. It can not only be suitable for mobile digital devices, i.e., phones, but also be useful for medical imaging devices, i.e., CT machines. Also, we will deal with image denoising with non-reference images in the future.

\section*{Acknowledgments}
This work was supported in part by National Natural Science Foundation of China under Grant 62201468, in part by the China Postdoctoral Science Foundation under Grant 2022TQ0259 and 2022M722599, in part by the Youth Science and Technology Talent Promotion Project of Jiangsu Association for Science and Technology under Grant JSTJ-2023-017.
%\section*{References}
%% References


% \bibliographystyle{elsarticle-num}
\bibliographystyle{elsarticle-harv}
\bibliography{references}
%\bibliography{sample}

%-----------------------------------------------------------------------------------------------------
\end{spacing}
\end{document}